\documentclass{sig-alternate-05-2015} \usepackage{tikz}
\usepackage{color}

\begin{document}

\setcopyright{acmcopyright}

\title{A Blueprint for Interoperable Blockchains}
\subtitle{[Vision Paper]}
%
%
%
%
%

\numberofauthors{3} 
%
\author{
%
%
\alignauthor
Tien Tuan Anh Dinh\\
       \affaddr{Singapore University of Technology and Design}\\
       \email{dinhtta@sutd.edu.sg}
\alignauthor
Anwitaman Datta\\
       \affaddr{Nanyang Technological University}\\
       \email{anwitaman@ntu.edu.sg}
\alignauthor
Beng Chin Ooi\\
       \affaddr{National University of Singapore}\\
       \email{ooibc@comp.nus.edu.sg}
}

\sloppy
\maketitle
\begin{abstract}
Research in blockchain systems has mainly focused on improving security and bridging the performance gaps
between blockchains and databases. Despite many promising results, we observe a worrying trend that the blockchain landscape is fragmented in which many systems exist in silos. Apart from a handful of general-purpose blockchains, such as
Ethereum or Hyperledger Fabric, there are hundreds of others designed for specific applications and typically do not talk to each other.  

In this paper, we describe our vision of interoperable blockchains. We argue that supporting interaction among
different blockchains requires overcoming challenges that go beyond data standardization. The underlying
problem is to allow smart contracts running in different blockchains to communicate. We discuss three open
problems: access control, general cross-chain transactions, and cross-chain communication. We describe
partial solutions to some of these problems in the literature. Finally, we propose a novel design to
overcome these challenges. 


\end{abstract}



\section{Introduction} 
A blockchain is a replicated, tamper-evident database designed for hostile environments. It consists of a
number of nodes some of which may be malicious (or Byzantine), and exposes an append-only log (or ledger)
abstraction. The ledger stores {\em transactions} that modify some global states. In the canonical example,
that is cryptocurrencies~\cite{btc_origin, eth_origin}, the global states are user accounts and native currencies,
and the ledger contains transactions transferring currencies from one account to another. Most 
blockchains support smart contracts which let users define their own states and codes that modify the
states. Smart contracts are stored in the ledger, which means they are replicated and kept consistent by all
the blockchain nodes. 

Early blockchains have poor performance and scalability~\cite{blockbench} compared to distributed databases,
and many smart contracts contain security flaws~\cite{dao,luu16}. Much research effort has thus been spent
on improving performance~\cite{sharding_19,omniledger} and hardening smart contracts~\cite{securify,zeus}, resulting in
a new generation of systems that achieve comparable performance to some databases, and smart contracts that
are much more secure. 

\begin{figure}
\centering
\includegraphics[width=0.4\textwidth]{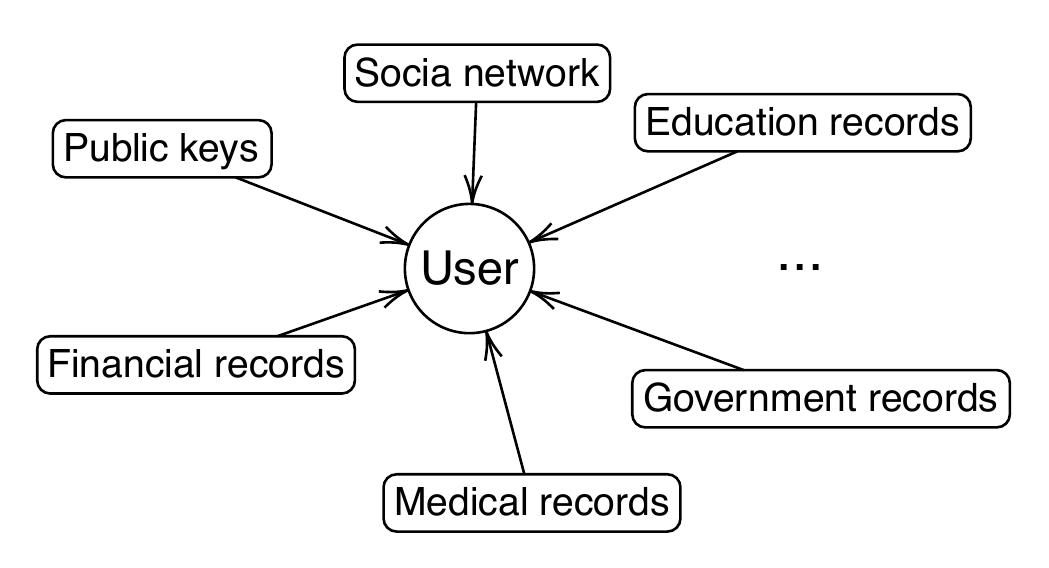}
\caption{Example of a fragmented blockchain landscape. There is one blockchain per aspect of user
identity. But these systems do not talk to each other.} 
\label{fig:id_silos}
\end{figure}

However, we observe that the blockchain ecosystem has a long tail. On the one hand, there are a small number
of hugely popular, general-purpose blockchains. One example is Ethereum~\cite{eth_origin}, a public (or
permissionless) blockchain in which anyone can join. Another is Hyperledger Fabric~\cite{hyperledger},
a private (or permissioned) blockchain in which nodes' identities are known to each other. On the other hand,
there are thousands of other blockchains designed for specific applications, most of which are only in early
stages of development. In particular, there are over $2500$
blockchains for cryptocurrencies\footnote{\url{https://www.coinlore.com/all_coins}}, $40$ for 
healthcare\footnote{\url{https://www.disruptordaily.com/blockchain-market-map-healthcare/}}, $100$ for
identities management\footnote{\url{https://github.com/peacekeeper/blockchain-identity}}, $25$ for
IoT\footnote{\url{https://www.disruptordaily.com/blockchain-market-map-iot/}}. More importantly, these
blockchains do not interoperate, i.e., they exist in silos. Figure~\ref{fig:id_silos} shows an example of
many blockchains storing different aspects of a user's identity (fragmentation across verticals), including public key~\cite{uport},
financial records~\cite{eth_origin}, medical records~\cite{medilot}, education certificates~\cite{blockcerts}
and government records~\cite{chromaway}. Inside each vertical, there are multiple isolated systems,
compounding further fragmentation. The user is overburdened with managing numerous credentials to access
different systems. Furthermore, the user is responsible for keeping data between different systems consistent,
for example, to update his financial records when a new government record is issued that certifies his tax
  exemptions.   

Like in a database, data fragmentation is a major source of inefficiency because it incurs management overheads. A
typical database defragmentation algorithm would move data around so that they are in the same place.
This solution, however, does not translate directly to blockchains, which would involve adding another
blockchain that aggregates all data in one place, which is unwieldy and goes against the very idea of decentralization.
Instead, we argue that a practical direction is to make existing blockchains interoperable. 
Interoperability entails more than standardizing message format across different blockchains. It requires carefully
designed protocols allowing one blockchain to {\em access} data of another. Blockchain interoperability has
received little attention from the research community.  Interledger~\cite{interledger} and Cosmo
network~\cite{cosmo} are the only two notable examples that aim at {\em connecting}
blockchains.\footnote{However, there are no complete implementations of these proposals.} The former focuses
on connecting payment networks, whereas the latter focuses on low-level message exchanges.  

In this vision paper, we go deeper than connecting blockchains, and seek designs that allow smart contracts in
one blockchain to access data of other smart contracts in another blockchain. To this end, we identify three
challenges. The first is to support secure, fine-grained access control for smart
contracts. Although any access policy can arguably be implemented and enforced inside a stand-alone smart contract, we note that
good security practice requires decoupling control policy and enforcement from actual data access. The second
challenge is to support general cross-chain transactions which are different to atomic
swaps~\cite{atomic_swap,decred} and cross-chain deals~\cite{deals}. This difference is similar to that  between
general and one-shot transactions in distributed databases~\cite{granola,hstore}. The final challenge is to
enable communication between smart contracts, which is currently not possible for contracts in 
different blockchains~\cite{oraclize,towncrier}.  

We propose to overcome the first challenge by decoupling access control from user smart contracts. We present a framework
consisting of a high-level language for specifying fine-grained policies, and a runtime environment with
access to historical states. For the second challenge, we design a protocol that supports a transaction model
similar to Sinfonia's mini-transaction~\cite{sinfonia} model which captures conditional cross-chain swaps.
For the third challenge, we propose a publish/subscribe framework that let smart contracts send and receive
messages to and from the outside world.    

In summary, our paper makes the following contributions:
\begin{itemize}
\item We highlight the problem of blockchain fragmentation, and propose to mitigate it by
making blockchains interoperable. 
\item We discuss three challenges in designing interoperable blockchains: access control, general
transactions, and cross-chain communication.  
\item We present high-level solutions to these challenges. 
\end{itemize}

Section~\ref{sec:challenges} presents the problem in detail by elaborating the three challenges. It
explains why current systems fail to adequately address these. Section~\ref{sec:approach}
sketches our solutions. We wrap up in Section~\ref{sec:conclusions} with concluding remarks.


\section{Challenges}
\label{sec:challenges}
We first motivate the problem of blockchain interoperability through an example of an auction application.
Next, we discuss three crucial challenges that need to be addressed. We describe how state-of-the-art systems
only provide partial building blocks for addressing these challenges. 

\begin{figure}
\includegraphics[width=0.48\textwidth]{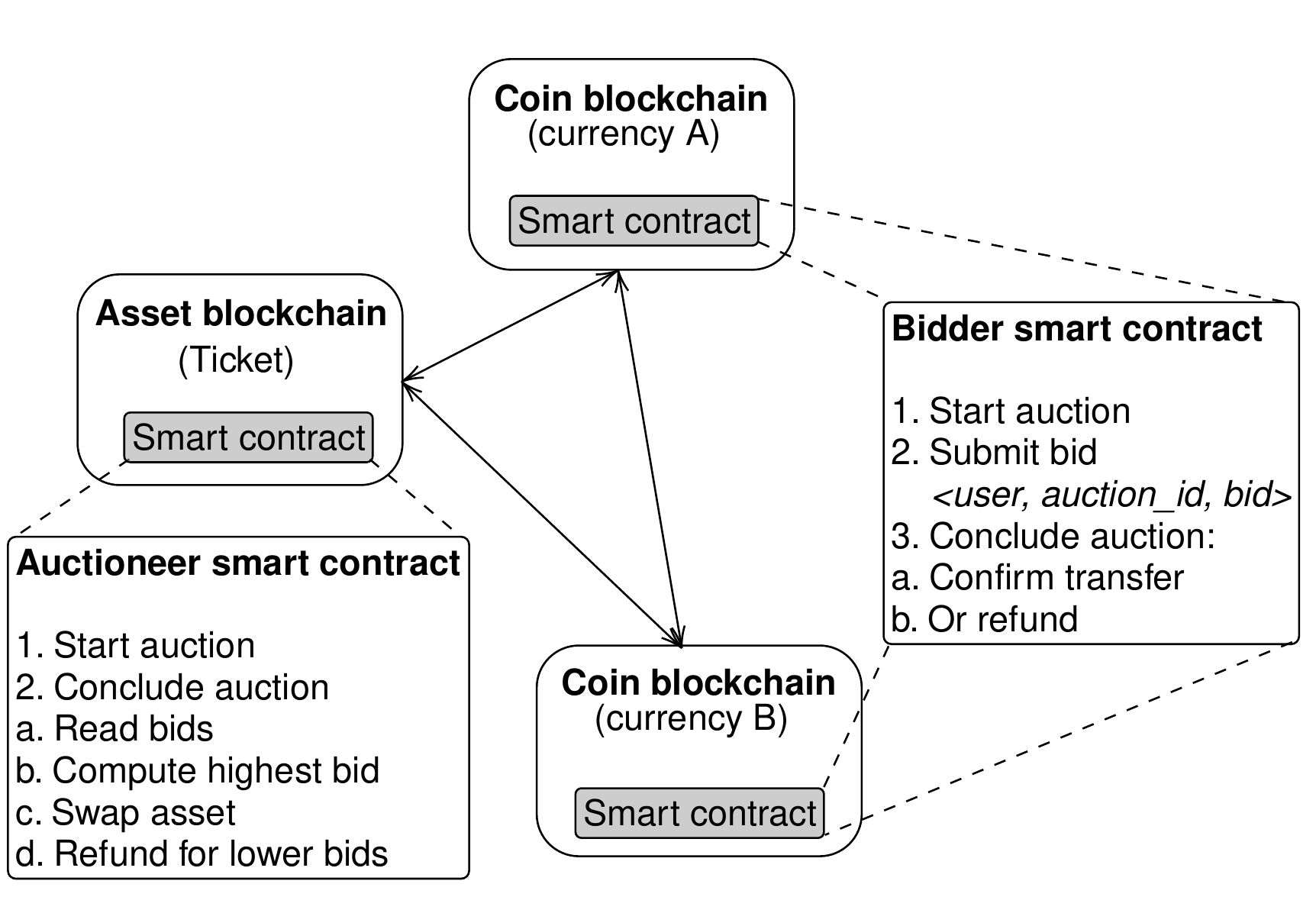}
\caption{Example of cross-chain, general transactions spanning three blockchains. The auction transaction
involves writes which are dependent on reads.}
\label{fig:example}
\end{figure}

\subsection{Motivating Example}
Alice owns a coveted football ticket stored on a ticket blockchain (many permissioned blockchains are designed
for asset management, and therefore can be used to implement such ticket application). Bob and
  Carol are avid football fans and both want to buy the ticket. They are users of the ticket blockchain, and
  each of them owns currencies at other coin blockchains (for example Ethereum or XRP~\cite{xrp}). Alice
  wants to start an auction on her ticket, to which Bob and Carol submit their bids. The ticket
  blockchain does not support bid submissions, thus Bob and Carol have to declare their bids in their own
  blockchains. This auction is implemented as smart contracts running on the three blockchains realizing a
  distributed workflow~\cite{workflowchain}.    

Figure~\ref{fig:example} shows the contracts' logics and how they interact. In particular, Alice
uses an Auctioneer contract that escrows the ticket when the auction starts. Bob and Carol specify their
bids to a Bidder contract that escrows the amount of currencies submitted. Before they can submit bids, the
Auctioneer contract must start the auction at the Bidder contract.  When Alice decides to end the
auction, the Auctioneer contract reads submissions from the Bidder contracts on the other two blockchains. It
then computes the highest bid, taking into account prevailing exchange rates among different
currencies. Finally, it atomically transfers ownership of the ticket to the highest bidder, deducts the bid
amount in corresponding blockchain, and releases the other for refund.  

There are three implications in enhancing existing blockchains to support the interaction above. First,
the Bidder contract must execute read queries from another contract. This warrants an access control mechanism
which allows Bidder to specify and enforce policies on who can access which piece of data. Second, the conclude transaction in
Auctioneer is a general, interactive transaction that involves cross-chain reads and cross-chain writes that
are dependent of the read values. Unlike one-shot transactions or atomic cross-chain swaps, this type of
transactions incurs more than two network round-trips to first execute and then commit the transaction. Third,
both Auctioneer and Bidder must be able to send and receive messages from each other. For instance, Auctioneer
must be able to send a read request and listen for the response value from Bidder. However, this 
is beyond the capability of current smart contracts. 

\subsection{Access Control}
One may question the need for access control, since one important property of blockchain is
transparency which means the blockchain nodes see all the data. We note that for permissionless blockchains,
only primitive forms of access control (all-or-nothing access based on possession of private keys) is
meaningful. However, for a permissioned blockchain there is a trust boundary between nodes inside and outside
of the system. Therefore, access control is needed to determine blockchain membership. Furthermore, the nodes
may wish to protect privacy of their data, therefore access from outside the system (be it
from a user or a smart contract), especially write access, must be restricted. For example, a blockchain for
financial transactions between major banks may only grant limited read access to auditors~\cite{zkledger}.  

Existing permissioned blockchains have built-in support for coarse-grained access policy. For instance,
Hyperledger Fabric provides a {\em membership service} that uses access control list (ACL) to determine who
can read or write to the ledger and the event streams. Fine-grained
policies, which give user more control over the data, are not supported. Examples of such
policies include the following:
\begin{itemize}
\item {\em Data-dependent policies:} only give access when some predicates over the current data are true. In the
auction example, the user can submit at most one bid (or write) to the Bidder contract only when the auction
is still ongoing.  

\item {\em Time-dependent policies:} access is restricted by time. In the context of blockchains, a time
duration may be represented as a range of blocks. In our example, the writing period for the Bidder contract
can be set to expire after a certain number of blocks. 

\item {\em Provenance-dependent policies:} only give access when some predicates over the data history (or
provenance) are true. In our example, the Bidder contract may only allow the Auctioneer contract to start an
auction if the latter has not started more than a certain number of auctions in the last $b$ blocks. Another policy may be to suspend a certain Auctioneer contract from starting an auction if it has accepted a bid from a corrupt user. 
 
\item {\em Aggregate policies:} only give access to summaries of the data. In the auction example, the Auctioneer
contract may grant read access to the total values of all winning bids over a certain period to an auditor.
Similarly, the Bidder contract may grant access to statistics of user bids to an external data analyst.
Further constraints such as the exposed statistics being differentially private may also be imposed. 

\end{itemize}
 
The first challenge is how to add support for these policies, particularly in a set-up where the information
is dispersed across different blockchains. We note that they may all be implemented explicitly inside the
smart contract if all the necessary information is self-contained in a blockchain, but even that has two major
limitations. The first limitation is related to performance. In particular, in most blockchains the contract
can only access the latest data. Thus, to support provenance-dependent policies the contract will need to
explicitly keep track of historical data, for example, the VersionKVStore contract in~\cite{blockbench}. This
introduces redundancy and unnecessary performance overhead because historical data is already stored in the
blockchain. The second limitation is related to software engineering. More specifically, smart contract
development is an error-prone process often carried out by non-security experts, therefore the risk of
implementing incorrect policies is high. Furthermore, many contracts share similar policies, and implementing
them in every contract implies duplication of effort. These problems are amplified when data from across
multiple blockchains are involved, due to concerns of correctness and consistency.

Another related challenge is the efficiency of enforcement. For example, aggregate and differentially private
policies require non-trivial computation to answer queries. Although smart contract execution is not the
performance bottleneck, long execution time may break the incentive mechanisms which
are important for security~\cite{verifier_dilemna}.  

In summary, the principal challenges for access control are: support for fine-grained policies, ease of
development, and enforcement efficiency.

\subsection{Cross-Chain Transactions}
Transaction is an important abstraction in modern distributed databases~\cite{spanner,calvin}.
Blockchains assume a much stronger failure model, namely Byzantine model, than the crash failure model in
databases. Consequently, supporting transactions in blockchains is more challenging. We consider only 
cross-chain transactions, instead of the normal, intra-chain transactions which are executed serially
by all nodes in the blockchain and therefore satisfying all of the ACID properties.

Read-only transactions, often considered the simplest type of transaction, must ensure that responses from the  
blockchain are correct {\em and} fresh. In our example, the Auctioneer issues a read query to the Bidder and
waits for the response. It needs to verify that the response has not been tampered with, and that it is
fresh (not a replayed message). Freshness can be achieved by including a nonce in the request, and
verifying it in the signatures of at least $f+1$ blockchain nodes.\footnote{$f$ is the fault-tolerance
threshold, usually expressed in terms of the number of Byzantine nodes.} For query correctness, we note that
blockchain nodes can execute read queries directly on top of the storage, or through the smart contract. The
former is applicable only when the values are mapped directly to the ledger data structure, for example block
information or key-value tuples. In this case, each node reads the values directly from the storage, then
includes integrity proofs as part of the response. However, complex smart contracts do not map their data
directly to the ledger data structure. For example, bid tuples $(\texttt{user, auction\_id, bid})$ are
stored in a list or a set as opposed to as key-value tuples. Therefore, in general case, read queries must be
executed through smart contracts, and the responses are signed by correct blockchain nodes, which ensures
correctness and authenticity.  However, the cost is significant because every query must go through consensus.
Reducing this cost is a major challenge. 

Concurrency arises due to cross-chain transactions. In our example, when the Auctioneer contract wants to
conclude an auction, it sends a read request to Bidder and waits for the response. The Bidder contract
executes the read query as a transaction, but it does not understand that the read is part of another, larger
transaction from another blockchain. Before Auctioneer finishes its transaction, the Bidder contract accepts
another bid that is the highest. Without coordination, the Auctioneer transaction is unaware of this new bid
and its execution becomes invalid. In other words, the transaction is not serializable. State-of-the-art
blockchain systems use lock based concurrency control to achieve serializability~\cite{sharding_19}. Specifically, smart contracts maintain one lock per key-value tuple, and a transaction must acquire all locks
before it can commit. This strategy is safe, but as demonstrated in~\cite{sharding_19}, it results in high abort
rate and significantly reduced throughput. 

Atomicity in the context of cross-chain transactions, like in traditional databases, means that the
transaction executes to completion or not at all. In our example, the transaction that concludes auction must
be atomic; otherwise it could so happen that the highest bidder gets the ticket without paying, or users who
submitted lower bids fail to get their refunds. The classic solution to atomicity is the two phase commit
(2PC) protocol, which is run by a transaction coordinator. Implementing 2PC in blockchain is more challenging
than in databases, because the coordinator is not trusted. Both~\cite{sharding_19,deals} propose to run
2PC in a Byzantine fault-tolerant network, i.e. in another blockchain. 

State-of-the-art atomic cross-chain transactions, namely~\cite{sharding_19,deals,atomic_swap},
consider only one-shot transactions~\cite{hstore}. Like in databases, this type of transaction does not
require communication between the blockchains during both the execution and commit phase. As a result, the
transaction can be executed and committed in two network round trips from the coordinator to the blockchains.
Atomic swaps and cross-chain deals~\cite{deals} involve only write operations to the blockchain, therefore
they are both one-shot. In particular, cross-chain deals are modeled as $n\times n$ write matrix for $n$ users. 

We argue that support for more general transactions is needed. In our example, the conclude transaction at the
Auctioneer contract is not one-shot, because the write operations are dependent on the read values. This type
of interactive transactions, like in databases, is expensive because they require multiple network round trips
to execute and commit. Therefore, the challenge is to design an efficient protocol for cross-chain
transactions that are more general than one-shot. We note that~\cite{deals} mentioned support for conditional
swaps, which are similar to our example above, but did not describe how it works.  

In summary, the challenges in supporting cross-chain transaction are how to provide efficient read
transactions, how to extract more concurrency, and how to reduce communication for general transactions. 
They are the same challenges the database community wrestles with~\cite{spanner,janus,rococo}.

\subsection{Communication}
One major limitation of current smart contracts is that they do not talk to the outside world. They can only
access resources on the blockchain such as the ledger and execution engine. The contract treats local states and user inputs
as ground truth, and makes no assumption about data from other sources. Some external services, such as
Oraclize~\cite{oraclize}, provide authentic data feeds, such that messages signed by them are accepted by the
contract. Towncrier~\cite{towncrier} and PDFS~\cite{pdfs} also address authenticity problem, but without
relying on trusted parties. The former leverages trusted hardware~\cite{sgx}, while the latter depends on
users to audit transparent logs. For real-world data, we believe authenticity (as opposed to trustworthiness)
is the best that can be achieved. On the other hand, data that comes from another blockchain can be
considered trustworthy, under the assumption that the blockchain as a whole is trusted. 

Our goal is to design an efficient communication infrastructure for sending and receiving messages between
blockchains. Interledger protocol (ILP)~\cite{interledger} and Cosmos' inter-blockchain communication
(IBC)~\cite{cosmo} are two recent proposals for connecting blockchains. ILP is specific to cryptocurrencies,
and its design is optimized for sending payments from one blockchain to another. It is based heavily on
payment channels~\cite{sprites}, and therefore does not generalize to other blockchain applications. IBC
proposes a more generic communication protocol which handles data transport, authentication and connection
reliability. Beside being in early stage of development, IBC requires intrusive changes to the blockchain
stack, as it needs to be deeply integrated with the state-machine component of the blockchain node. Another
limitation of IBC is that it is a networking-layer protocol, therefore it leaves much complexity to the
application layer (i.e., the smart contracts). For example, IBC only relays messages and does not distinguish
between ones carrying data versus carrying proof. Finally, IBC is connection-oriented, which means both
communication endpoints have to maintain connection states. This connection model is suitable when the
endpoints are responsive, for example in a client-server model, and when the network is often reliable.
However, it is not suitable for blockchains, because the network is hostile (it may
have Byzantine nodes) and transaction latency is high.  
 
In summary, the challenge is to design a stateless communication protocol that ensures timely and
reliable delivery of messages. The protocol needs to work at the application layer, which understands and contains
optimizations for common requirements of cross-chain communication. Finally, the protocol needs to be resilient to network failures.  

\section{Possible Solutions}
\label{sec:approach}

\begin{figure}
\centering
\includegraphics[width=0.43\textwidth]{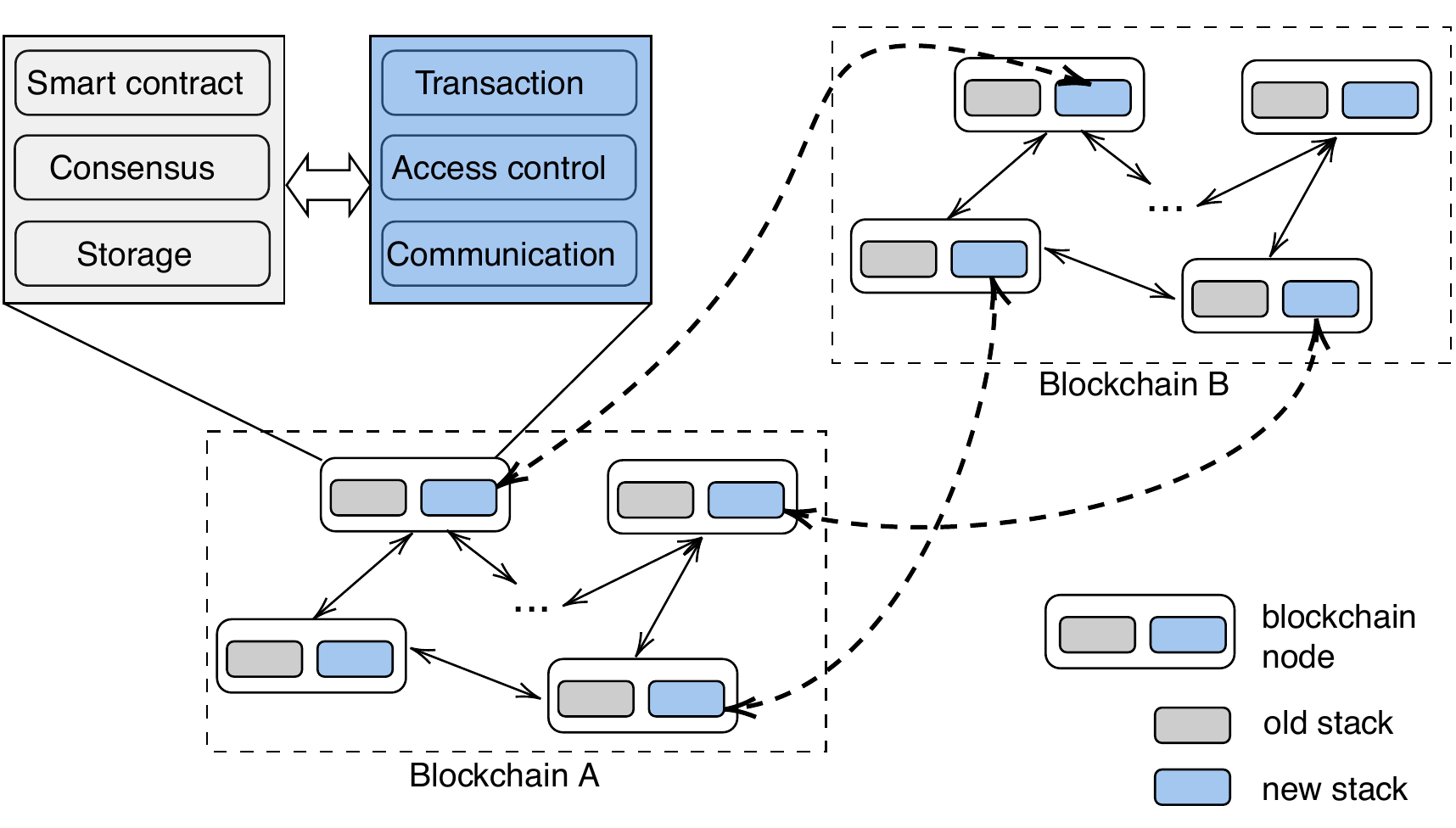}
\caption{The architecture for interoperable blockchains, with newly added components for transaction, access control and
communication. The old software stack remains unchanged.} 
\label{fig:architecture}
\end{figure}
Figure~\ref{fig:architecture} shows our proposed architecture for interoperable blockchains. The current blockchain stack,
consisting of a storage, consensus and smart contract components~\cite{blockbench}, remains unchanged. We add
new components that implement support for access control, transactions, and communication. This architecture
avoids designing a new blockchain from scratch, and allows for incremental adoption. In the following, we
sketch our high-level design for each of the new components. We describe how we expect to overcome the challenges
discussed in the previous section. We stress that our approach builds on existing solutions, and it is only to
start the discussion. More research is needed to find better solutions and to validate them. 

\begin{figure}
\centering
\includegraphics[width=0.35\textwidth]{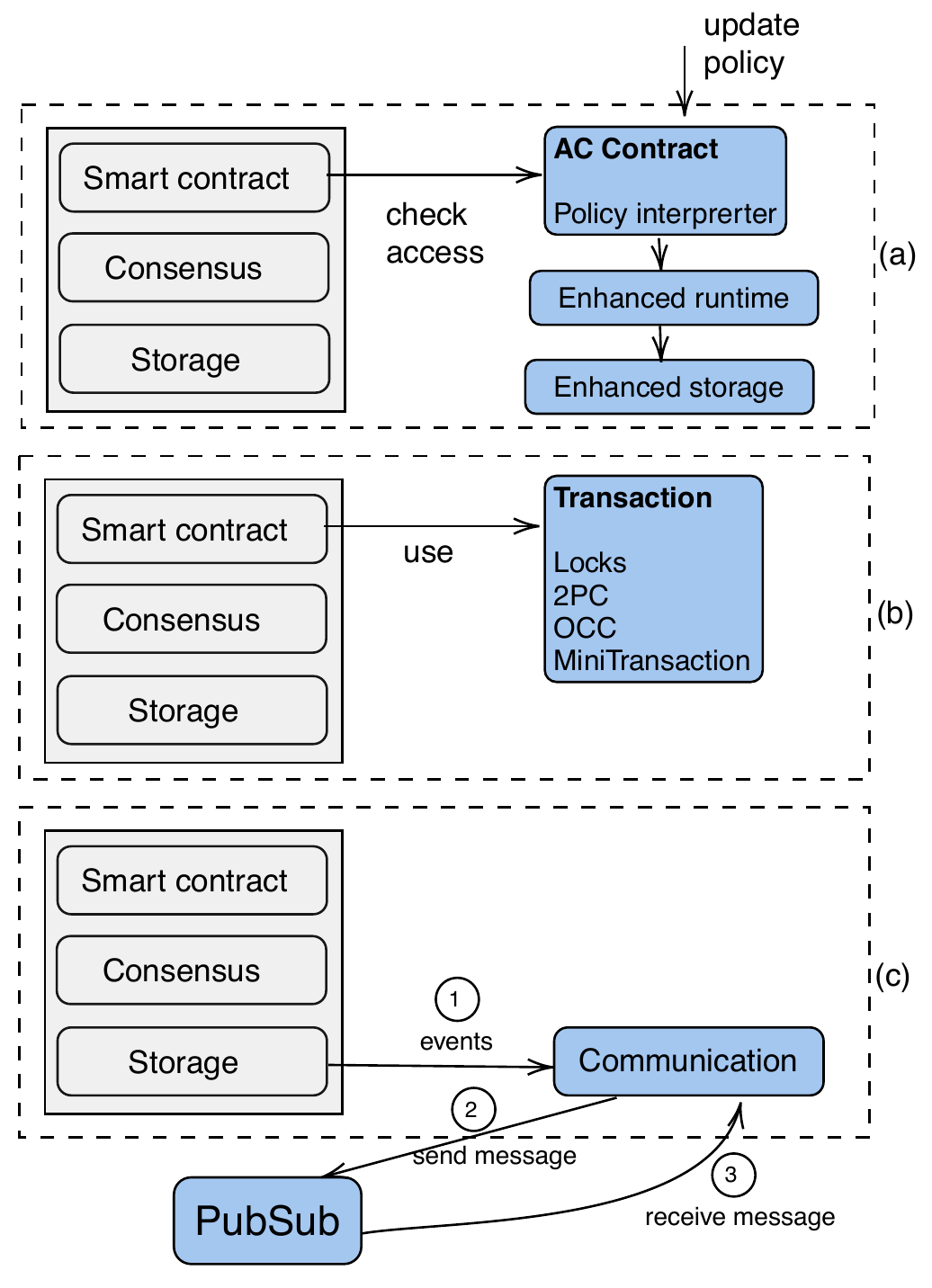}
\caption{Main components of access control, transactions and communication.}
\label{fig:design}
\end{figure}

\subsection{Access Control}
We decouple access control from user smart contracts and implement it as a {\em system contract}. As shown in
Figure~\ref{fig:design}[a], the user contracts can invoke the access control mechanism by calling another
contract within the same blockchain. We implement it as a contract as opposed to a system library because it
allows users to update policies without changing the software at all the nodes. This design solves the ease-of-development
challenge. 

To support fine-grained policies, we propose to design a policy language that is expressive and easy to use. This
language is declarative and can be based on either Google Firestore~\cite{firestore} or Datalog~\cite{datalog}. The
access control contract contains the language interpreter to parse and then enforce the policy. To support
access to historical data during enforcement, we propose to use LineageChain~\cite{lineagechain}, a
state-of-the blockchain storage providing rich data access to smart contracts. LineageChain and its
underlying engine called Forkbase~\cite{forkbase} have been evaluated on Hyperledger Fabric. More work is
needed to port them to Ethereum runtime. Finally,
to improve enforcement efficiency, we propose to add concurrency to the smart contract execution
engine~\cite{concurrency}. 
The fact that current blockchains execute transactions sequentially means there is plenty of room for improvement. 

\subsection{Cross-Chain Transactions}
Figure~\ref{fig:design}[b] shows the key components for transactions implemented as a system library
which a smart contract can invoke. To improve efficiency for read-only transactions, it is necessary to avoid
going through consensus. This means the storage must be able to produce integrity proofs for the read
values. We propose to enhance blockchain storage with more expressive, high-level data structures that are
verifiable and can be directly used by the smart contract. Examples include map and list structure.
Trillian~\cite{trillian} provides strong security properties for map structures. Forkbase~\cite{forkbase}
provides weaker guarantees for both map and list structure. Our proposed solution is based on Forkbase.  

To extract more concurrency from cross-chain transactions, we propose to add support for optimistic
concurrency control (lock-based mechanisms are still provided, because they work well under high contention).
The OCC protocol works the same way as in databases: it keeps track of read/write sets and aborts when there
are newer versions. We directly use the version tracking feature provided by the storage~\cite{forkbase} to
implement this. 

Finally, to reduce transaction communications, we propose to implement the mini-transaction
abstraction~\cite{sinfonia} which captures conditional data swaps. A mini-transaction comprises a compare, a
read, and a write phase with known read and write sets.  Although it does not capture all general, interactive
transactions, its execution and commit phase can be optimized to finish in two network round trips. For other unsupported
types of transactions, we provide general locks and 2PC implementation. During execution, all
locks are acquired. During commit, the Byzantine-tolerant 2PC is run by one of the blockchains. 

\subsection{Communication}
Figure~\ref{fig:design}[c] shows our communication infrastructure which is stateless, efficient, and works at the
application level. It is based on publish-subscribe system. A communication library runs at every blockchain
node and subscribes to the event streams produced by the blockchain stack. It then publishes the events to the
pub-sub system, which delivers them to the destination blockchain. A cross-chain event contains names of the
source, destination blockchain, and identifier of both the source and destination contract. 

The communication library automatically adds unique nonces to every sent event to guarantee freshness. We
also propose running a dedicated gateway node to which all nodes forward their events. This gateway then waits
and batches $f+1$ signatures of the same event before publishing them to the pub-sub system, which
significantly reduces network cost. 

We note that the pub-sub system is not a centralized component. One way to realize this system is to have the
communication blockchains running their own P2P pub-sub system. However, a more practical approach is to use
a third-party service such as Amazon SNS. This service does not have to be trusted for security, because
messages between blockchains are cryptographically signed. Furthermore, multiple services from different
providers can be used at the same time, without coordination among them, to increase quality of service and
avoid denial of service attacks (such as dropping of messages).  


\section{Concluding Remarks}
\label{sec:conclusions}

We have discussed related work extensively in the previous sections. Here, we highlight three orthogonal research
directions that may benefit from our work. First, off-chain scaling systems such as Plasma~\cite{plasma} and sidechains~\cite{sidechain} use multiple blockchains to improve overall throughput. Our work can be applied to
make communication between the sub-chains more efficient and richer. Second, there is a trend towards
decentralized systems, of which blockchain is only one example. Other examples include identity
management~\cite{coniks}, personal storage~\cite{upspin}, and social network~\cite{decent}. These systems
exist in silos, and our work can be applied to build novel applications on top of them. Finally, our approach can help realize general purpose distributed workflows \cite{workflowchain}.

We presented our vision of making blockchains interoperable, as a solution to the current fragmented blockchain
ecosystem. We discussed three challenges in realizing our vision: fine-grained access control, cross-chain
transactions, and cross-chain communication. Finally, we proposed solutions to address these challenges, which
serves as starting points for future research.

%
\bibliographystyle{acm}
{\scriptsize
\bibliography{ref}  
}
%
%
\end{document}